\documentclass[aip,jap,reprint]{revtex4-1}

\usepackage{graphicx,amsmath,amsfonts,amssymb,textcmds}

\begin{document}

\title{Diffusion-Emission Theory of Photon Enhanced Thermionic Emission Solar Energy Harvesters}
\author{Aapo Varpula}
\email{aapo.varpula@vtt.fi}
\author{Mika Prunnila}
\affiliation{VTT Technical Research Centre of Finland, P. O. Box 1000, FI-02044 VTT, Espoo, Finland}
\date{\today}

\begin{abstract}
Numerical and semi-analytical models are presented for photon-enhanced-thermionic-emission (PETE) devices. The models take diffusion of electrons, inhomogeneous photogeneration, and bulk and surface recombination into account. The efficiencies of PETE devices with silicon cathodes are calculated. Our model predicts significantly different electron affinity and temperature dependence for the device than the earlier model based on a rate-equation description of the cathode. We show that surface recombination can reduce the efficiency below 10 \% at the cathode temperature of 800 K and the concentration of 1000 suns, but operating the device at high injection levels can increase the efficiency to 15 \%.
\end{abstract}

\maketitle

\section{Introduction}

Currently solar energy is converted to electric power using two technologies: photovoltaic (PV) solar cells and concentrated solar power systems based on heat engines \cite{book:Kalogirou}. The former system requires low and the latter high operating temperatures. This discrepancy poses a challenge for combination of the two systems in tandem, where the heat engines exploit the waste heat of the PV system. A photon-enhanced-thermionic-emission (PETE) device proposed by Schwede et al. \cite{art:Schwede} is a PV device which benefits from high operation temperatures. It can be coupled to a heat engine, thereby allowing total efficiencies above 50\% to be potentially reached.

\begin{figure}[tbp]
\includegraphics[width=240pt]{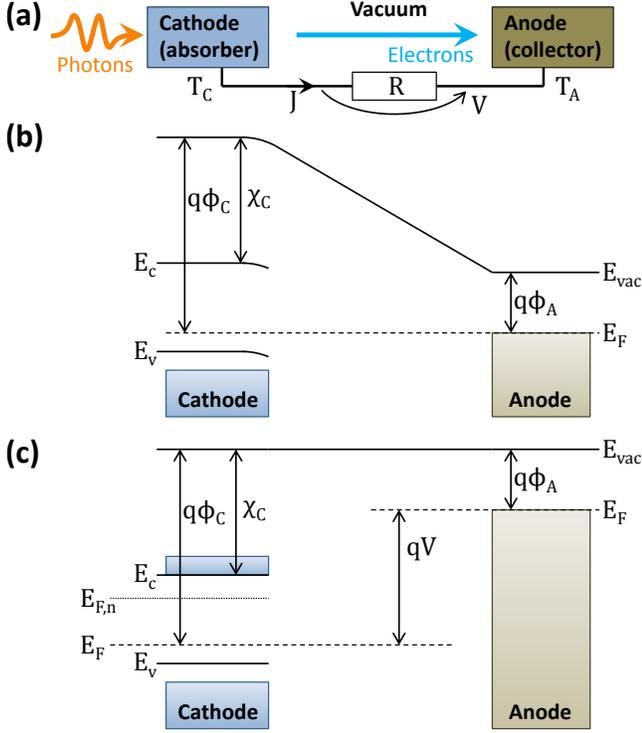}
\caption{(a) Schematic picture of the PETE device. $T_C$ is the absolute temperature of cathode, $T_A$ the absolute temperature of anode, $V$ the output voltage, $J$ the density of the output current, and $R$ a resistor representing the external load. The energy bands of the device (b) in the thermodynamical equilibrium and (b) in the flat-band case, which often yields the highest power output. $E_F$ is the Fermi level, $E_{F,n}$ is the quasi Fermi level of electrons, $q$ is the elementary charge, $\chi_C$ is the electron affinity of the cathode, $\phi_C$ and $\phi_A$ are the work functions of the cathode and the anode, $E_v$ is the valence band maximum, $E_c$ is the conduction band minimum, and $E_{vac}$ is the vacuum energy level. Here the anode is metal, but it can be an N-type semiconductor as well.}
\label{fig:PETEdevice}
\end{figure}

The PETE device is depicted in Fig.~\ref{fig:PETEdevice}. The photons are absorbed in the cathode, i.e. the absorber, which is a P-type semiconductor. The cathode material should have a suitably low energy gap so that most of the solar photons are absorbed. The absorbed photons are themionically emitted from the cathode to vacuum, where they travel to the anode, i.e. the electron collector. The surface of the cathode should have a low electron affinity in order to have reasonably strong thermionic emission. Electron affinity can be tuned significantly below the bulk value by different surface coatings (see Refs. \onlinecite{art:Schwede, art:Guo} and references therein). Even negative electron affinities can be obtained for silicon \cite{art:Guo}. These coatings, however, might not be stable at the high operation temperatures of PETE devices. The anode material can be metal or N-type semiconductor with suitably low work function in order to have high output voltage $V$ for the device.

In the PETE device model of Ref. \onlinecite{art:Schwede} the cathode material is described by a single rate equation that neglects many important effects, such as diffusion and realistic recombination. In this article, we present a model that takes the most of the relevant effects in semiconductors into account. We solve the electron density in the cathode numerically in the general case and derive also a semi-analytical model, which can be used at low injection levels. Silicon is a good candidate for PETE due to its rather low band gap, good thermal stability, high availability, cost-efficiency, and good manufacturability. Therefore, we use our models to calculate the characteristics of a PETE device with a silicon cathode using experimentally verified material data. We find that at low temperatures our model predicts a significantly lower efficiency than the model presented in Ref. \onlinecite{art:Schwede}. The overall temperature and electron affinity dependency of the efficiency differs also from that of Ref. \onlinecite{art:Schwede}. Furthermore, we show that surface recombination can reduce the efficiency of the PETE device below 10 \%, but operating the device at high injection levels can provide an enhancement where efficiency of 15 \% is reached.

\section{Theory}

\subsection{Semiconductor material model}

The Fermi level $E_{F}$ is solved numerically using the electroneutrality condition \cite{book:Sze} $n_{eq}+N_{A}^{-}=p_{eq}+N_{D}^{+}$, where $n_{eq}$ and $p_{eq}$ are the densities of electrons and holes in the thermodynamical equilibrium, and $N_{A}^{-}$ and $N_{D}^{+}$ are the densities of ionized acceptors and donors, respectively. For the temperature dependence of the band gap of silicon we use the standard formula \cite{book:Sze}
\begin{equation}
E_{g}(T) = E_g(0) - \frac{\alpha_g T^2}{T + \beta_g},
\end{equation}
where $E_g(0)=1.170$~eV, $\alpha_g=4.73 \times 10^{-4}$~eV$/$K, and $\beta_g = 636$~K.

The total minority-electron lifetime in bulk can be written as \cite{art:Altermatt_rev, book:Schroder, art:Altermatt}
\begin{equation}
\tau_n = \frac{1}{  B(n + p_{eq} ) +(C_n n + C_p p )(n + p_{eq}) + \frac{1}{\tau_{SRH}}},
\label{eq:tau_tot_app}
\end{equation}
where $B$ is the radiative recombination coefficient,  $n$ and $p$ are the densities of electrons and holes, respectively, $C_n$ and $C_{p}$ are the Auger recombination coefficients for electrons and holes, respectively, and $\tau _{SRH}$ is the Shockley-Read-Hall (SRH) recombination lifetime, which is directly proportional to the density of the deep-level impurities in the material, and it increases when the injection level is increased \cite{book:Schroder}. The SRH lifetime has constant saturation values both in low and high injection regimes \cite{book:Schroder}. Therefore, we use constant $\tau _{SRH} = 2.5$~$\mu $s to represent a high quality silicon \cite{art:Altermatt} at either low or high injection level. For silicon \cite{book:Schroder} $B=4.73\times 10^{-15}$~cm$^{3}/$s and $C_{p}=10^{-31}$~cm$^{6}/$s, which is a value measured at low injection level. In a P-type material $C_p$ determines the Auger lifetime in low and the ambipolar Auger coefficient $C_A = C_n + C_p$ in high injection conditions \cite{book:Schroder}, respectively. The experimental high-injection value for $C_A$ differs from $C_n + C_p$ measured at low injection levels \cite{art:Altermatt, art:Altermatt_rev}. We take the both injection regimes into account using an artificial value of $10^{-30}$~cm$^{6}/$s for $C_n$. For electron mobility $\mu_n$ in silicon we use a model \cite{art:Reggiani} optimized for a wide temperature range. Mobility is linked to the diffusion coefficient $D_n=k_{B}T\mu_n/q$, where $T$ is the absolute temperature and $q$ the elementary charge. For the absorption coefficient $\alpha$ of silicon we use a widely-used semi-empirical model \cite{art:Rajkanan, book:Green}.

\begin{figure}[tbp]
\includegraphics[width=240pt]{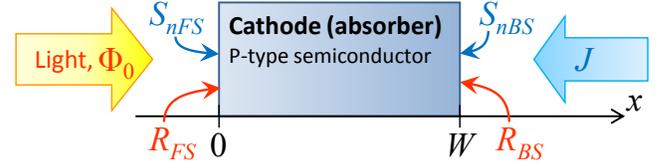}
\caption{Schematic picture of the cathode of the PETE device.}
\label{fig:PETEdevice_absorber}
\end{figure}

\subsection{Electron density}

The electron density $n$ in the cathode sketched in Fig.~\ref{fig:PETEdevice_absorber} can be calculated using the continuity equation, the generation and recombination rate equations, and the drift--diffusion current formulas for the charge carriers \cite{book:Sze}. We assume that the electric field inside the cathode is approximately zero, because the most of the voltage difference is across the vacuum gap and the photogeneration of electron-hole pairs supports the local approximate charge neutrality. This simplifies the solution considerably, since holes need not to be taken into account explicitly. In addition, the band bending near the cathode surface (see Fig.~\ref{fig:PETEdevice}b) is neglected, since the effect is small and the PETE device is mostly operated near the flat-band case (shown in Fig.~\ref{fig:PETEdevice}c), where the band bending disappears. The excess-electron density $\Delta n(x)=n(x)-n_{eq}$ can be described by
\begin{equation}
\frac{\mathrm{d}^{2}\Delta n}{\mathrm{d}x^{2}}=\frac{\Delta n}{L_{n}^{2}}-%
\frac{G_{n}(x)\tau _{n}}{L_{n}^{2}},  \label{eq:n_diff_cont_excess}
\end{equation}%
where $L_n = \sqrt{D_n \tau_n}$ is the diffusion length of electrons and the generation is given by
\begin{equation}
G(x)=\alpha \Phi _{0}\left( 1-R_{FS}\right) \left[ e^{-\alpha x}+R_{BS}e^{\alpha \left( x-2W\right) }\right] ,  \label{eq:G_exp_BSlight}
\end{equation}
where $\alpha$ is the absorption coefficient, $\Phi_0$ is the incident flux of photons, $R_{FS(BS)}$ is the front (back) surface reflection coefficient, and $W$ is the thickness of the cathode. The boundary conditions are
\begin{align}
\left. \frac{\mathrm{d}\Delta n}{\mathrm{d}x}\right\vert _{x=0}& =\frac{%
S_{nFS}}{D_{n}}\Delta n(0)  \label{eq:JnBC_n_front_Sn} \\
\left. \frac{\mathrm{d}\Delta n}{\mathrm{d}x}\right\vert _{x=W}& =-\frac{%
S_{nBS}}{D_{n}}\Delta n(W)-\frac{J}{qD_{n}},  \label{eq:JnBC_n_back_Sn}
\end{align}%
where $S_{nFS(BS)}$ is the front (back) surface recombination velocity and $J$ is the density of the output current.

The analytical solution exist when $\tau_n$ is constant, which is valid at low injection levels (i.e. $\Delta n \ll p_{eq}$). Assuming the low-injection case and using Eqs. \eqref{eq:JnBC_n_front_Sn} and \eqref{eq:JnBC_n_back_Sn} the solution of Eq. \eqref{eq:n_diff_cont_excess} at $x=W$ is given by
\begin{equation}
\Delta n(W)=-\frac{L_{n}k}{qD_{n}}J + \Delta n_{\mathrm{sun}},
\label{eq:n_linear}
\end{equation}%
where 
\begin{align}
k & = \frac{1}{a}\left[ \cosh {\left( \frac{W}{L_{n}}\right) }+\frac{L_{n}S_{nFS}}{D_{n}}\sinh {\left( \frac{W}{L_{n}}\right) }\right], \label{eq:k_JnBC_Sn} \\
a & = \sinh{\left( \frac{W}{L_n} \right)} + \frac{L_{n}S_{nFS}}{D_n}\cosh{\left( \frac{W}{L_n} \right)}+\frac{bL_{n}S_{nBS}}{D_n}, \\
b & = \cosh{\left( \frac{W}{L_n} \right)} + \frac{L_{n}S_{nFS}}{D_n} \sinh{\left( \frac{W}{L_n} \right)},
\end{align}
and
\begin{align}
& \Delta n_{\mathrm{sun}}=\int\limits_{0}^{\infty }\mathrm{d}\lambda \;\frac{%
\tau _{n}\alpha \Phi _{0}\left( 1-R_{FS}\right) }{1-\alpha ^{2}L_{n}^{2}}%
\left( e^{-\alpha W}\left( 1+R_{BS}\right) \phantom{ \frac{A_A}{A_A} }%
\right.   \notag \\
& +\frac{L_{n}}{a}\left\{ -\frac{S_{nFS}}{D_{n}}-\alpha -R_{BS}e^{-2\alpha
W}\left( \frac{S_{nFS}}{D_{n}}-\alpha \right) \right.   \notag \\
& \left. \left. +be^{-\alpha W}\left[ -\frac{S_{nBS}}{D_{n}}+\alpha
-R_{BS}\left( \frac{S_{nBS}}{D_{n}}+\alpha \right) \right] \right\} \right),
\label{eq:b_BSlight_hyp}
\end{align}
where $\lambda$ is the wavelength of photons. The first term in Eq.~\eqref{eq:n_linear} describes the fact that the electrons need to diffuse from bulk of the cathode to the back surface in order to be emitted. When current is drawn, $\Delta n(W)$ will decrease because of this diffusion process. The second term in Eq.~\eqref{eq:n_linear} represents the excess electrons due to the photogeneration.

\subsection{Output current}

The density of the cathode current, which corresponds to electrons emitted from the cathode, can be written using the quasi Fermi level $E_{F,n} = E_F + k_B T_C \ln{( n/n_{eq} )}$ as
\begin{equation}
J_C=A_{C}^{\ast }T_{C}^{2} \exp{\left( -\frac{ \Delta E_C }{k_{B}T_C} \right)} \frac{n}{n_{eq}},
\label{eq:JC}
\end{equation}
where $\Delta E_C = q\phi_C$ when $V \leq V_{fb}$, and $\Delta E_C = q\phi_C + q (V-V_{fb})$, when $V > V_{fb}$, $A_{C}^{\ast }$ is Richardson's constant, and $T_C$ is the absolute temperature of the cathode. The flat-band voltage $V_{fb}$ is defined as
\begin{equation}
V_{fb}=\phi _C-\phi _A,
\end{equation}
where $\phi _{A}$ is the work function of the anode material and
\begin{equation}
\phi _C = \frac{1}{q} \left( E_{c}-E_{F}+\chi_C \right)
\end{equation}
is the work function and $\chi _C$ the electron affinity of the cathode material. The density of the anode current, which corresponds to electrons emitted from the anode, is given by 
\begin{equation}
J_{A}=A_{A}^{\ast }T_{A}^{2}\exp {\left( -\frac{\Delta E_A }{k_{B}T_{A}} \right) },  \label{eq:JA}
\end{equation}
where $\Delta E_A = q\phi_A  + q(V_{fb} - V)$ when $V \leq V_{fb}$, and $\Delta E_A = q\phi_A$, when $V > V_{fb}$, $A_{A}^{\ast }$ is Richardson's constant of the anode, and $T_{A}$ is the absolute temperature of the anode. We assume that all the electrons emitted from the cathode are collected by the anode and vice versa:
\begin{equation}
J=J_{C}-J_{A}.  \label{eq:Jtot}
\end{equation}

The flat band voltage $V_{fb}$ is an important parameter for the efficiency of the PETE device. At voltages $V$ above $V_{fb}$ the additional energy barrier $q(V-V_{fb})$ for the electrons emitting from the cathode appears (see Eq.~\eqref{eq:JC}). This decreases $J_C$ considerably. At voltages $V$ below $V_{fb}$ the electrons emitted from the anode are hindered by the energy barrier $q(V_{fb}-V)$ (see Eq.~\eqref{eq:JA}). This allows $J_A$ to be reduced by lowering $V$. In general, the highest output power will often be obtained near $V=V_{fb}$. However, when $T_C \gg T_A$, the high thermal energy of the cathode electrons allows the range $V > V_{fb}$ to be used as well.

Using $n_{eq} = N_c \exp{[ -(E_c-E_F)/(k_BT_C) ]}$ and Eqs.~\eqref{eq:n_linear} and \eqref{eq:JC}--\eqref{eq:Jtot} the output electric current density can be written alternatively as 
\begin{equation}
J = J_\mathrm{sun} - J_\mathrm{dark},
\label{eq:Jtot_final}
\end{equation}
where the photocurrent is given by 
\begin{equation}
J_\mathrm{sun} = \frac{q D_n}{L_n d} \: \Delta n_\mathrm{sun},
\label{eq:Jsun}
\end{equation}
where 
\begin{equation}
d = \begin{cases} 
\frac{q D_n N_c}{L_n A^\ast_C T^2_C} \exp{\left( \frac{\chi_C}{k_BT_C} \right)} + k & \text{for } V \leq V_{fb} \\
\frac{q D_n N_c}{L_n A^\ast_C T^2_C} \exp{\left[ \frac{\chi_C + q \left( V - V_{fb} \right) }{k_BT_C} \right]} + k & \text{for } V > V_{fb}.
\end{cases}
\label{eq:Jtot_denom}
\end{equation}
The dark current is given by
\begin{equation}
J_\mathrm{dark} = \frac{q D_n n_{eq}}{dL_n} \left\{ \frac{A^\ast_A T^2_A}{A^\ast_C T^2_C} \exp{\left( \frac{q\Delta E_\mathrm{dark}}{k_B T_C} \right)} - 1 \right\} 
\label{eq:Jdark}
\end{equation}
where $\Delta E_\mathrm{dark} = \phi_C + ( V - \phi_C ) T_C/T_A$ when $V \leq V_{fb}$, and $\Delta E_\mathrm{dark} = V + \phi_A( 1 - T_C/T_A )$, when $V > V_{fb}$. The dark current is the electric current that flows through the device when there is no illumination. Under illumination it usually decreases the total current, thus it should be eliminated. This can be done by choosing $V$ optimally. However, if $T_A$ is very low, the direction of $J_\mathrm{dark}$ can change, and the device harvests energy also from the thermal energy of cathode electrons similarly as a thermionic converter.

Eqs.~\eqref{eq:Jsun} and \eqref{eq:Jtot_denom} show that in order to maximize $J_\mathrm{sun}$ $\chi_C$ and $k$ should be as small. On the contrary, then $J_\mathrm{dark}$  will also increase. $J_\mathrm{sun}$ can be increased and $J_\mathrm{dark}$ decreased by increasing $T_C$. $J_\mathrm{dark}$ can be reversed by having high $\phi _A$ and low $\phi _C$ (this will reduce $V_{fb}$). In addition, $T_{A}$ and $A_{A}^{\ast }$ should be small and $T_{C}$ and $A_{C}^{\ast }$ should be large.

\section{Results and discussion}

The semi-analytical model, Eqs.~\eqref{eq:k_JnBC_Sn}, \eqref{eq:b_BSlight_hyp}, and \eqref{eq:Jtot_final}--\eqref{eq:Jdark}, applies in the low injection condition ($\Delta n \ll p_{eq}$) with a P-type cathode material ($p_{eq} \gg n_{eq}$) \cite{book:Schroder}. The numerical calculations were performed using Eqs. \eqref{eq:n_diff_cont_excess}--\eqref{eq:JnBC_n_back_Sn} and \eqref{eq:JC}--\eqref{eq:Jtot}.  In both models $V$ was optimized numerically for maximum output power. The AM1.5 direct+circumsolar spectrum applicable to solar concentrators was used in the calculations. We use {\textlangle}111{\textrangle} silicon as the cathode surface, for which \cite{book:Sze} $A_C^\ast = 264$~A$/$cm$^2/$K$^2$. For the material parameters of the anode we use the same values as Schwede et al. \cite{art:Schwede}, $\phi_A = 0.9$~V and $A_A^\ast = 120$~A$/$cm$^2/$K$^2$. $T_A$ was set at 573.15~K in order the heat engine potentially coupled to the anode to have a reasonably high efficiency \cite{art:NH3_H20_Rankine, art:Solar_polygeneration}. We also assume $R_{FS}=0$ and $R_{BS}=1$ for simplicity. The effect of $R_{BS}$, however, is not very large (see below), since it has an effect only on the absorption of low energy photons.

\begin{figure}
\includegraphics[width=240pt]{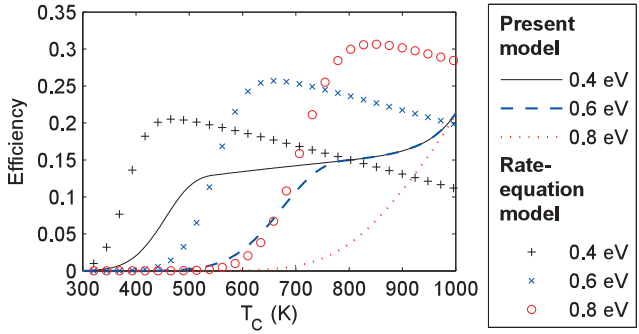}
\caption{Efficiencies of PETE devices with silicon cathodes with various electron affinities $\chi_C$ as functions of the cathode temperature $T_C$ calculated using the numerical model with $N_A = 10^{18}$~cm$^{-3}$ and $W = 5$~$\mu$m and without surface recombination at the concentration of 1000 suns. The results from the rate-equation model \cite{art:Schwede} were calculated with $E_g(T)=1.12$~eV. }
\label{fig:Eff_Chi}
\end{figure}

The efficiencies of PETE devices with various values of $\chi_C$ are plotted as functions of $T_C$ in Fig.~\ref{fig:Eff_Chi}. They increase with increasing $T_C$ and decreasing $\chi_C$ due to the increase of $J_C$. Using $R_{BS}=0$ instead reduces the efficiency from 13.0~\% to 11.7~\% at 550~K in the case $\chi_C=0.4$.

Fig.~\ref{fig:Eff_Chi} shows also the efficiencies given by the simple rate-equation model \cite{art:Schwede}, which assumes that all photons with energy greater than $E_g$ are uniformly absorbed in the cathode. Only the uniform radiative recombination is taken into account with a general model based on the black-body radiation. At low $T_C$ the rate-equation model suggests much higher efficiencies than the present more complete model. This is mostly due to the facts that all the possible photons are absorbed and the Auger recombination is not included in the rate-equation model.

The efficiency is a balance between many opposing effects, which depend on $N_A$ and $W$: Increasing $N_A$ increases $V_{fb}$ which increases the efficiency. On the other hand, high $N_A$ reduces the efficiency due to the Auger recombination which is proportional to $N_A^2$. The effect of the bulk recombination can be reduced by decreasing $W$, but then some of the photons will not be absorbed. Although 5~$\mu$m is already small thickness for a silicon absorber, but, if $W$ is increased to 50~$\mu$m while keeping $N_A$ constant, the efficiency reduces from 13.0~\% to 8.4~\% at 550~K in the case $\chi_C=0.4$~eV. At high $T_C$ the efficiency curves of the present model unite regardless of the differences in $\chi_C$. The reason for this is that the total current does not depend on $\chi_C$ in this range: When $T_C \gg T_A$ the range $V > V_{fb}$ can be utilized and the output current density can be written using Eqs.~\eqref{eq:JC}--\eqref{eq:Jtot} as $J = A_C^\ast T_C^2$exp$[(E_{F,n}-E_F - qV - q\phi_A)/(k_BT_C)] - A_A^\ast T_A^2 $exp$(- q\phi_A/k_BT_A)$. The rate-equation model does not behave like this at high $T_C$, because $\Delta n$ decreases much faster with increasing $T_C$ than in the present model. This is caused mostly by the differences in the modeling of recombination and the narrowing of the band gap with increasing $T_C$ which is taken into account only in the present model. The latter effect enhances both photogeneration and thermal generation of electrons, which causes the apparent efficiency increase at $T_C>800$~K.

\begin{figure}
\includegraphics[width=240pt]{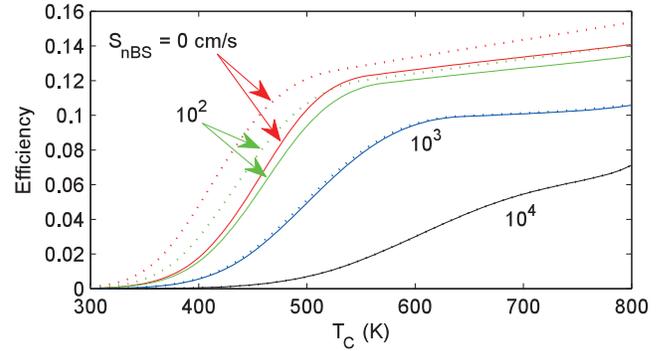}
\caption{Efficiency of a PETE device with a silicon cathode and various back-surface recombination velocities $S_{nBS}$ as function of the cathode temperature $T_C$ calculated using the numerical (solid lines) and semi-analytical (dotted lines) models with $S_{nFS}=10^2$~cm/s, $\chi_C = 0.4$~eV, $N_A = 10^{18}$~cm$^{-3}$, and $W = 5$~$\mu$m at the concentration of 1000 suns.}
\label{fig:Eff_Sn}
\end{figure}

The effect of $S_{nBS}$ on the efficiency of the PETE device is shown in Fig.~\ref{fig:Eff_Sn}: Even a low value of $S_{nBS}$ reduces the efficiency remarkably. Similar results were also obtained with various values of $S_{nFS}$. The efficiency is more sensitive to $S_{nFS}$ than $S_{nBS}$ because photogeneration is much stronger near the front surface than near the back surface. The semi-analytical model is in a very good agreement with the numerical model when the recombination velocities are $\geq 1000$~cm/s since the injection level is below 0.1 at $T_C > 550$~K. In the low-injection conditions achieving surface recombination velocities less than 100~cm/s requires usually use of back surface field structures \cite{book:Sze}. However, the surface recombination velocity decreases rapidly when the injection level is increased \cite{book:Sze} and values below 2~cm/s can easily be reached at injection level of unity (details depend on the properties of the surface states) \cite{book:Schroder}. Therefore, the case of zero surface recombination can actually be realistic for a PETE device.

\section{Conclusions}

In summary, we have built a theoretical model for PETE devices. Our model takes electron diffusion, inhomogeneous photogeneration, and bulk and surface recombination into account. In comparison to the rate equation model of Ref.~\onlinecite{art:Schwede} our model predicts different dependency of efficiency on such paramaters as cathode electron affinity and temperature. In most cases our model also predicts lower efficiency. Especially, the surface recombination present on real surfaces can reduce the efficiency to extremely low values. However, the surface recombination might have only a very weak effect on the performance, since the PETE device works often in the high-injection conditions where the surface recombination can be rather small \cite{book:Schroder}. We finally point out that the realization of the PETE device requires choices of many parameter values and materials which should all be optimized. In addition, the heat balance between the PETE device and the heat engine coupled to the anode, which was not considered in this article, should be managed as well. Full optimization with heat balance modelling will be left for future studies. 

\begin{acknowledgments}
Fruitful discussions with J. Tervo, J. Ahopelto, K. Reck, O. Hansen, and P. Kuivalainen are gratefully acknowledged. This work has been financially supported by Nordic Energy Research (project HEISEC) and by the Academy of Finland (grant nr. 252598).
\end{acknowledgments}



\end{document}